\documentclass[final,5p,times]{amsci}

\usepackage{amssymb,amsfonts,amsmath,amstext,amsgen,amsopn,amsxtra,indentfirst,lscape}

\journal{Volume 101, Number 3, Pages 184--187 (May/June 2013 Issue)}

\begin{document}

\begin{frontmatter}

\title{Why Does Nature Form Exoplanets Easily?}

\author{Kevin Heng}
\ead[csh]{kevin.heng@csh.unibe.ch}
\address{University of Bern \\ Center for Space and Habitability \\ Sidlerstrasse 5, CH-3012, Bern, Switzerland}

\begin{abstract}
\textit{The ubiquity of worlds beyond our Solar System confounds us.\vspace{0.05in}\\
{\scriptsize Kevin Heng received his M.S. and Ph.D. in astrophysics from the Joint Institute for Laboratory Astrophysics (JILA) and the University of Colorado at Boulder. He joined the Institute for Advanced Study in Princeton from 2007 to 2010, first as a Member and later as the Frank \& Peggy Taplin Member. From 2010 to 2012 he was a Zwicky Prize Fellow at ETH Z\"{u}rich (the Swiss Federal Institute of Technology). In 2013, he joined the Center for Space and Habitability (CSH) at the University of Bern, Switzerland, as a tenure-track assistant professor, where he leads the Exoplanets and Exoclimes Group. He has worked on, and maintains, a broad range of interests in astrophysics: shocks, extrasolar asteroid belts, planet formation, fluid dynamics, brown dwarfs and exoplanets. He coordinates the Exoclimes Simulation Platform (ESP), an open-source set of theoretical tools designed for studying the basic physics and chemistry of exoplanetary atmospheres and climates (www.exoclime.org). He is involved in the CHEOPS (Characterizing Exoplanet Satellite) space telescope, a mission approved by the European Space Agency (ESA) and led by Switzerland. He spends a fair amount of time humbly learning the lessons gleaned from studying the Earth and Solar System planets, as related to him by atmospheric, climate and planetary scientists. He received a Sigma Xi Grant-in-Aid of Research in 2006.} \vspace{0.05in}\\
Edited by Fenella Saunders.  \texttt{www.americanscientist.org }
} 
\end{abstract}

\end{frontmatter}

An area of research that has attracted a lot of attention in the field of astronomy and astrophysics is planet formation: the study of how planets (in our Solar System) and exoplanets (orbiting other stars) form. Astronomers harness the power of telescopes with meter-sized or larger mirrors to search the night sky for exoplanets---and they find loads of them. In the past two years, NASA's Kepler Space Telescope has located nearly 3,000 exoplanet candidates ranging from sub-Earth-sized minions to gas giants that dwarf our own Jupiter. Their densities range from that of styrofoam to iron. Astronomers find them close to their parent or host stars with scorching temperatures of a few thousand degrees; they also find them distant from their stars, more than 10 times farther away than Jupiter is from our Sun. Perhaps most significant, the Kepler results demonstrate that rocky exoplanets are common in our local cosmic neighborhood---and by extension, our universe at large. Nature seems to have a penchant for forming exoplanets.

As astrophysicists, our goal is to construct hypotheses to explain what we see in nature. We create model universes and exoplanetary systems on paper and in our computers. When our hypotheses stand the tests of data and time, they eventually become accepted as theories. Hypotheses of planet formation are usually forged within two accepted paradigms: core accretion and gravitational instability. Core accretion is the ``bottom-up" approach: Large objects form from smaller ones, eventually building up to exoplanets. Gravitational instability is the ``top-down" method: Exoplanets form directly from larger structures in the primordial disks of gas and dust orbiting young stars. But when astrophysicists zoom in on the physical details, we find ourselves (and our hypotheses) flummoxed and, quite simply, outclassed by nature. Dust grains do not seem to readily stick. Even if rocks form, they then drift into the star much too quickly, fast enough to preclude their coalescence into larger objects. These larger, kilometer-sized objects, known as planetesimals, are in principle the building blocks of planets. For our Solar System, theorists struggle in modeling to form the rocky cores of the gas giants, Jupiter and Saturn, before the primordial gas of the natal disk dissipates. Even forming Neptune within the paradigm of core accretion takes too long due to its relative remoteness from the Sun. The devil is in the details and, unfortunately, they do matter when trying to construct synthetic exoplanets.

\vspace{0.15in}
\noindent
\textbf{Rocks are Hard to See}
\vspace{0.05in}

\begin{figure*}
\begin{center}
\includegraphics[width=1.5\columnwidth]{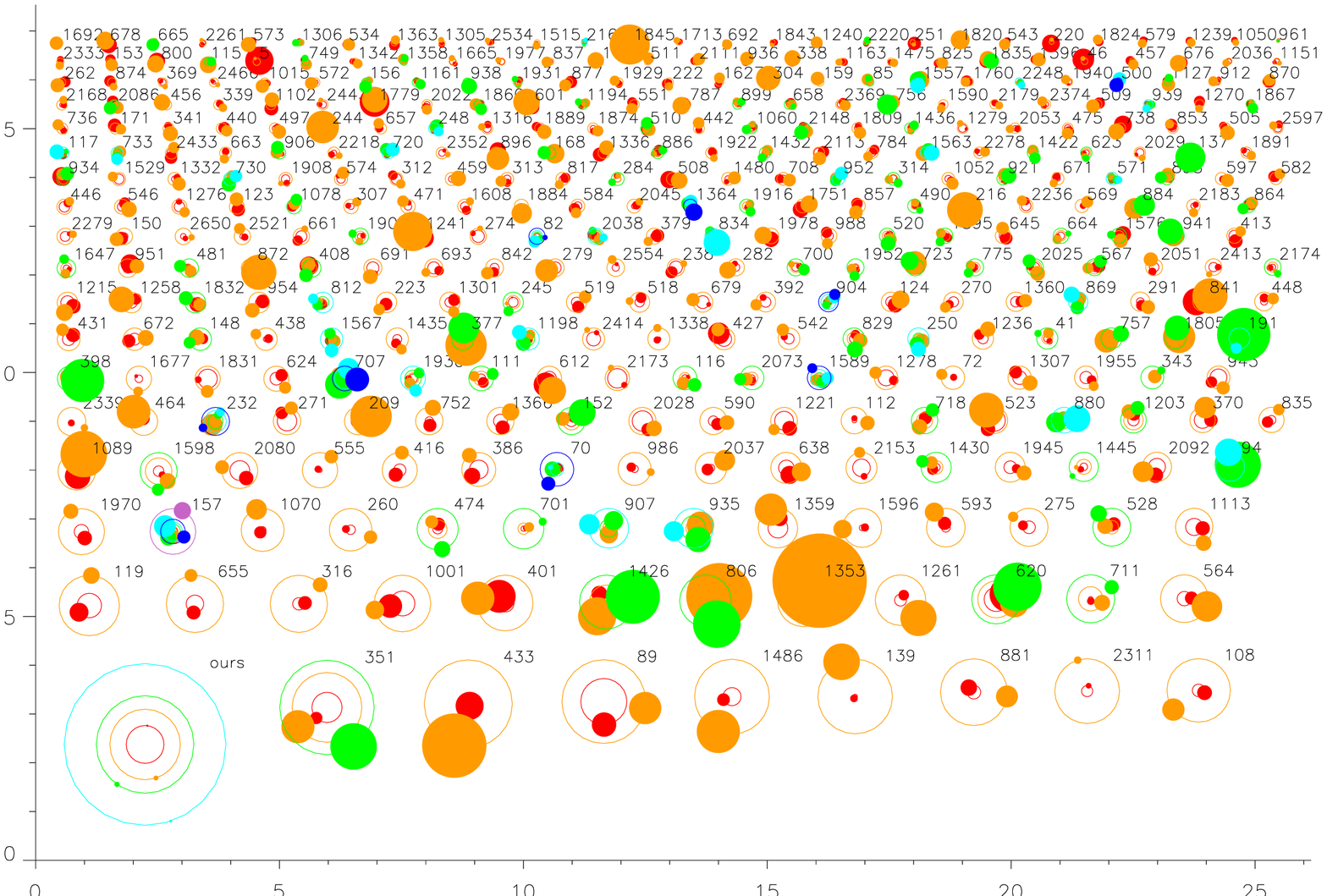}
\end{center}
\vspace{0.2in}
\caption{NASA's Kepler Space Telescope mission has found thousands of candidate exoplanets that are passing in front of their stars in its field of view.  Of these, 361 systems of more than one planet are depicted here, along with their catalog numbers (a separate color is used for each planet in a system). Many of these systems have multiple exoplanets tightly packed together. The orbital distances are on one scale, and the planet radii are on a separate scale. The terrestrial planets of the Solar System (at lower left and labeled ``ours") are also shown for comparison. This
conglomeration is called the Kepler Orrery, named after the mechanical device that shows the positions and motions of the planets and moons in the Solar System. (Image courtesy of Dan Fabrycky of the University of Chicago.)}
\end{figure*}

The hardest thing to observe with a telescope is a rock. Astronomers can see disks of gas and dust swirling around nascent stars, as well as fully formed exoplanets orbiting mature stars. But it is terribly difficult---if not impossible---to detect the presence of planetesimals, at least outside of our Solar System. An entire cottage industry of astronomers is engaged in the study of protostellar or protoplanetary disks, the purported birthplaces of exoplanets. They find a rich cacophony of features in these disks: asymmetries, gaps, warps, dust made from exotic substances. Sometimes, they even find exoplanets embedded in the disks (which often cause the features themselves). A decisive test of planet formation hypotheses is to study the properties of planetesimals within these disks, but these objects are practically invisible. Planetesimals are neither numerous (compared to dust grains) nor large enough (compared to exoplanets) to re-emit significant radiation that can be detected by telescopes. Any link to them is indirect and uncertain at best, made through assumptions of their relationship to the observed dust grains. Even estimating the masses of these disks remains a crude exercise at best and requires Solar SystemÐcentric assumptions on the opacity of the dust grains and the relative amounts of dust and gas present in the disk. We find ourselves bombarded with information, but knowledge eludes us.

The prevalence of rocky exoplanets in nature lends credence to the paradigm of core accretion. Astrophysicists hoping to model formation by this mechanism appear to have their work cut out for them: They must start with micrometer-sized dust grains and congregate them to produce larger particles, either on paper, in computer simulations or in the laboratory. The task is daunting, because there is conceivably a factor of a billion in size between dust grains and planetesimals. The protoplanetary disk conspires to exacerbate matters---gas orbiting a star tends to move at slower speeds than for the dust grains, because of the self-exerted pressure support.  The dust grains experience a ``head wind," which acts like friction, causing them to spiral into the star. The time scale on which this ``gas drag" phenomenon occurs is, uncomfortably, shorter than any conceivable time scale for grain growth. This conundrum is commonly called the ``meter-size problem," because it afflicts meter-sized objects the most when they are located at the same distance from the Sun as the Earth.

Proposed solutions to the meter-size problem are plentiful. Some astrophysicists devote their careers to re-creating in their computers the conditions for encounters between dust grains. These simulations capture the intricate details of collisions, both constructive and destructive: coagulating, chipping, bouncing, deflecting, shattering. Dust grains of all sizes and shapes are studied. Again, we are awash in information, but knowledge is slower to come. Nature is hinting to us that planet formation is a robust phenomenon---surely, the mechanism involved cannot be privy to all of the micro-details of the dust grains and must be related somehow to the global properties of the protoplanetary disk. Although it cannot serve as a proof, we are guided by principles such as Occam's razor---when faced with many explanations, we pick the simplest one.

\vspace{0.15in}
\noindent
\textbf{Working on Theories}
\vspace{0.05in}

Theorists studying core accretion use two approaches. The first is to isolate a piece of the puzzle---such as the meter-size problem---and study the microphysics involved. Such approaches shed light on important pieces of the puzzle of planet formation but lose the big picture. A complementary approach is to use population synthesis, which is an attempt to incorporate all of the physics and chemistry involved in core accretion, starting from dust grains and ending up with exoplanets. Gaps in our knowledge of the details currently prevent the population synthesis framework from being a complete theory, but its redeeming quality is that it provides some falsifiable predictions.

Another solution is to bypass the intermediate steps of growth between dust grains and planetesimals. Protoplanetary disks are usually massive enough that gravity may triumph over pressure support. The fragments that result from gravitational instability may be the size of planetesimals or even exoplanets. The issue then is not that one cannot form structures, but rather that the gravitational instability paradigm lacks predictive power. Again, the devil is in the details. A successful theory of gravitational instability should start from the initial properties of the protoplanetary disk and the conditions imposed upon it by the star---the strength of irradiation, the metallicity of the natal gas (the abundance of elements that are heavier than hydrogen and helium), the mass of the disk---and predict the number of exoplanets that ensue, including their masses and interior structures. In other words, theorists need to build a population synthesis framework for gravitational instability. A theory with such completeness currently eludes us.

The true answer may lie in between the two paradigms. Giant structures may form in the disk, which then collect dust grains within them to form larger particles. An intriguing possibility is the existence of vortices, essentially giant cyclones or hurricanes forming out of the gas in the disk. Vortices have the ability to trap particles within them, much like dust devils on Earth---an old concept with its roots spanning back to the German philosopher Immanuel Kant. An attractive feature of this idea is that vortices would arise naturally from the turbulent gas if it behaves like a two-dimensional fluid. A parcel of fluid ``lives" in a two-dimensional world if it is sufficiently buoyant in the third dimension---as it is displaced from its plane of existence, it returns to its original position quickly. The gas swirling around a protoplanetary disk may be regarded as a fluid; the disk also has a finite thickness. Dust tends to collect mostly at the midplane of the disk. It turns out that the midplanes of protoplanetary disks behave like three-dimensional fluids, whereas locations farther away from the midplane, where the solid material needed to grow structure is found in less abundance, behave like two-dimensional fluids. A successful theory of planet formation involving vortices has to identify a mechanism for producing them, predict their lifetimes and elucidate the means by which they will be destroyed.

Perhaps an easier way to provide constraints on hypotheses of planet formation is simply to stare at the planets themselves. Gas giants appear to be more common around stars with higher metallicities, consistent with the notion that one needs more solid material to construct larger cores and trigger runaway accretion of the natal gas. No trend in stellar metallicity is found for the occurrence of rocky exoplanets. When they are found, they tend to be social creatures, located in systems with other rocky brethren. These exoplanetary systems also tend to be ``flat," just like our Solar System---the orbits of the rocky exoplanets lie roughly within the same plane. We now know that ``hot Jupiters"---gas giants found implausibly close to their host stars, such that their temperatures are a few thousand degrees---are oddballs (comprising less than one percent of the entire exoplanet population) and loners (without nearby exoplanets as companions), despite being the most common type of exoplanets initially found due to their brightness and large sizes. Furthermore, hot Jupiters are often found in orbital planes that are misaligned with the spin axes of their host stars---in stark contrast to the systems hosting multiple, rocky exoplanets---perhaps providing a clue that they were delivered to their present locations via some kind of scattering mechanism.

\vspace{0.15in}
\noindent
\textbf{Taking a Look}
\vspace{0.05in}

Some of the Kepler telescope discoveries present invaluable puzzles and challenges to our current ideas of planet formation. The Kepler-11 system plays host to six exoplanets with radii ranging from twice to five times that of Earth. Three of these exoplanets have densities less than that of water. All six appear to lie roughly in the same orbital plane. The Kepler-16 system consists of a pair of stars orbited by a Saturnlike exoplanet, a harsh environment for any exoplanet to survive in because of the enhanced gravitational tugs of the stars. A diminutive red dwarf sits at the center of the Kepler-32 system, yet it is orbited by five exoplanets within a distance a third the size of Mercury's orbit. Perhaps the most puzzling case study comes from the Kepler-36 system. Two planets are found at roughly the same distance from the star: one with a density less than that of water, whereas the other is as dense as iron. Theorists are just coming to grips with these discoveries. The venerable theory of migration---the notion that exoplanets and their progenitors drift through the gaseous disk during assembly---is coming under scrutiny. It may turn out that migration, although an attractive idea for constructing our Solar System, is not really needed to form exoplanetary systems populated by close-in super Earths---Earthlike exoplanets somewhat larger than our Earth, which seem to be an omnipresent breed. Our Solar System appears not to be the dominant outcome of planet formation and it may be exerting a provincial influence on our theoretical ideas. In seeking to map out the universe, we find ourselves to be the exception rather than the norm.

An intriguing idea, suggested during a recent conference on Kepler exoplanets by Renu Malhotra of the University of Arizona, is to search for the transits of planetesimals around the corpses of stars known as white dwarfs. During a transit, a body passes in front of its host star, causing a dip in electromagnetic emissions from the star. By coincidence, the relative size of a planetesimal orbiting a white dwarf is equivalent to that of an Earth orbiting a Sunlike star. Because the Kepler Space Telescope is detecting Earthlike exoplanets around Sunlike stars, the reasoning is that it will be able to do the same for planetesimals transiting white dwarfs. If this idea meets with success, it will be the first time information on planetesimals outside of the Solar System will be obtained. Another fascinating idea concerns the hunt for moons around exoplanets---``exomoons." The idea is that these exomoons are formed during the final stages of assembly, during a phase of planet formation known as ``clean-up," and may provide some clues on the properties of planetesimals.

If nothing else, one upshot from the numerous recent results is that planet formation is hardly a straightforward process, perhaps going some distance to explaining why nature forms planets and exoplanets more easily than theorists are capable of doing. But ultimately, we still seek a theory of planet formation because we wish to determine how common exoplanets, and life, are in the universe. The Kepler Space Telescope is showing us that about one in 10 stars may have an exoplanet that resembles Earth. For the aficionados, this frequency factor of 0.1 is termed ``eta Earth"; it is one of the several factors in Drake's equation, which seeks to quantify how common exoplanets are around stars, how many of these exoplanets are capable of harboring life, and how many of them actually do host life. From extrapolating the Kepler results, Earthlike exoplanets likely number in the billions for our galaxy alone; unless life is an exceedingly rare event, it must exist elsewhere in the universe. We may never answer all of these questions, but understanding planet formation is a step in the right direction.

\vspace{0.15in}
\noindent
\textbf{Acknowledgment}
\vspace{0.05in}

The author acknowledges financial and logistical support from the University of Bern, the University of Z\"{u}rich and the Swiss-based MERAC Foundation. He is grateful to Eric Agol, Dan Fabrycky and Jack Lissauer for enlightening conversations during the Kepler multi-planet conference, held at the Aspen Center for Physics in February 2013, Scott Tremaine and Lucio Mayer for constructive comments on an earlier version of the article, and George Lake for encouragement.

\label{lastpage}

\end{document}